\documentclass{sig-alternate}

\usepackage{subfigure,amsmath,hyperref}
\usepackage{url}
\newcommand{\sech}{\text{sech}}

\begin{document}

\conferenceinfo{}{}
\title{
Numerical Solutions to the Sine-Gordon Equation
}
%% \subtitle{ %% extended abstract
%% \titlenote{ %% blah
%% }
%% }

\numberofauthors{1}

\author{
\alignauthor
Paul Rigge \\
\affaddr{Department of Electrical Engineering and Computer Science} \\
\affaddr{University of Michigan} \\
\affaddr{Ann Arbor, MI 48109} \\
\email{riggep@umich.edu}
}

\maketitle

\begin{abstract}
The sine-Gordon equation is a nonlinear partial differential equation.
It is known that the sine-Gordon has soliton solutions in the 1D and 2D cases, but such solutions are not known to exist in the 3D case.
Several numerical solutions to the 1D, 2D, and 3D sine-Gordon equation are presented and comments are given on the nature of the solutions.
\end{abstract}

%%\terms{}

%%\keywords{}

\section{Introduction}
Many problems in science and engineering can be expressed as a partial differential equation (PDE).
Finding accurate numerical solutions is often important for these applications, but there are many challenges associated with computing these solutions.
Nonlinear PDEs can cause numerical methods to destabilize\cite{AsMc04}.
Finding accurate solutions can require high resolution, large amounts of memory, small time-steps, and a lot of computation time.

Numerical methods that compute solutions for PDEs must approximate derivatives. One way to achieve this is with a finite difference; one of the simplest such methods is the forward difference given by
\begin{equation}
  u_x \approx \frac{u(x+\delta x) - u(x)}{\delta x}
\end{equation}
\cite[Ch. 1]{Tre00} shows that the finite difference approximation will converge to the true value at a rate polynomial in $\delta x$; that is, as $\delta x$ gets smaller, the approximation gets better at a polynomial rate. Getting a very accurate approximation of the derivative may require $\delta x$ to be extremely small, which may exhaust memory requirement or take a lot of time, especially in the 3D case. \cite[Ch. 4]{Tre00} goes on to show that spectral methods converge faster than any polynomial, so we can get very accurate approximations with a relatively larger $\delta x$. This can save a lot of time and memory. If periodic boundary conditions are used, we use the Fourier transform to compute approximate spatial derivatives.

This paper considers the sine-Gordon equations in 1, 2, and 3 dimensions. The 1D problems considered here are all suited to being run on one processor. However, as the dimensionality is increased, several issues make using multiple processors necessary. Computing multi-dimensional Fourier transforms amounts to computing multiple one-dimensional Fourier transforms. These one-dimensional transforms are independent and can be computed more quickly in parallel. Memory requirements also increase dramatically as the dimensionality increases. Problem sizes of interest, especially in 3D, may require more memory than one machine can reasonably have, so using multiple computers may be necessary. Using multiple computers also gives more total cache, which generally improves performance.

\section{Background}
\subsection{Sine-Gordon Equation}
The sine-Gordon equation is a nonlinear partial differential equation given by
\begin{equation}
  u_{tt}-\Delta u = -\sin u \label{eq:sg}
\end{equation}
This equation arises in many different applications, including propagation of magnetic flux on Josephson junctions, sound propagation in a crystal lattice, and several others discussed in \cite{ScChMc73}. We consider the 1, 2, and 3 dimensional cases given by
\begin{subequations}
\begin{align}
  &u_{tt}-u_{xx}&= -\sin u \label{eq:sg1d} \\
  &u_{tt}-u_{xx}-u_{yy}&=-\sin u \label{eq:sg2d} \\
  &u_{tt}-u_{xx}-u_{yy}-u_{zz}&=-\sin u \label{eq:sg3d}
\end{align}
\end{subequations}

The one dimensional sine-Gordon equation is exactly integrable. This means a large class of exact solutions can be found using the inverse scattering transform, including a special class of solutions called solitons\cite{Ga67}. Solitons are nonlinear effects that lead to travelling waves that are localized and permanent. Solitons are interesting physically and as a way to check numerical codes. For the sine-Gordon equation, some of the explicit exact solutions cannot be easily evaluated\cite{MaTl06}. For these solutions, it may be may be easier to compute the solution numerically and can still give great insight. The sine-Gordon equation is known to have soliton solutions in the 1 and 2 dimensional cases\cite{GrMa78}.

The sine-Gordon equation has a Hamiltonian $E$ given by
\begin{subequations}
\begin{align}
E_K&= \int \frac{1}{2}u_t^2 \mathrm{d}\mathbf{x}\\
E_S&= \int \frac{1}{2}|\nabla u|^2 \mathrm{d}\mathbf{x}\\
E_P&=\int 1-\cos u \mathrm{d}\mathbf{x}\\
E&=E_K+E_S+E_P
\end{align}
\end{subequations}
We give each component of the Hamiltonian a name and interpretation\cite{Ra01}. $E_K$ is the kinetic energy, $E_S$ is the strain energy, and $E_P$ is the potential energy. $E$ is the total energy and is constant.
These quantities are useful in evaluating the accuracy of numerical schemes and the nature of a solution.

\subsection{Discrete Fourier Transform}
The Discrete Fourier Transform (DFT) can be used to approximate derivatives. The DFT and inverse DFT are given by
\begin{align}
\hat{v}_k&=h\sum_{j=1}^N e^{-ikx_j}v_j,\; k=-\frac{N}{2}+1,\ldots,\frac{N}{2} \\
v[j]&=\frac{1}{2\pi}\sum_{k=-N/2+1}^{N/2} \hat{v}_ke^{ikx_j},\; j=1,\ldots,N
\end{align}

Conceptually, we can think of the DFT as decomposing $v[j]$ into into a set of orthogonal complex exponentials. Each $\hat{v}_j$ represents the magnitude and phase of its corresponding complex exponential.

For smooth functions, the Fourier coefficients converge faster than any polynomial\cite{Tre00}. Using the Fast Fourier Transform (FFT) algorithm, the DFT can be computed in $O(n \log n)$ time\cite{CoTu65}.

Computing the derivative in Fourier space is a simple multiplication by the wave number $k$ and $i$.
\begin{align}
  \frac{\partial v[j]}{\partial x} &= \frac{1}{2\pi}\sum_{k=-N/2+1}^{N/2} \frac{\partial}{\partial x}\hat{v}_ke^{ikx_j},\; j=1,\ldots,N \\
&= \frac{1}{2\pi}\sum_{k=-N/2+1}^{N/2} ik\hat{v}_ke^{ikx_j},\; j=1,\ldots,N
\end{align}

\subsection{Numerical Method}
Following \cite{DoSc10} and \cite{CloMuiRig12}, we use a leapfrog method to approximate the second derivative in time with a central difference given by:
\begin{align}
u_{tt} &\approx \frac{u^{n+1}-2u^n+u^{n-1}}{\delta t^2} \label{eq:leapfrogtime}
\end{align}
where $\delta t$ is the time step size and $u^n$ is the approximation to the function $u$ at the $n$th time step.

We approximate the spatial derivative with a spectral method. Differentiation in real space becomes multiplication by the wave number in Fourier space, giving
\begin{align}
  \Delta u &= \left(\frac{\partial^2}{\partial x^2}+\frac{\partial^2}{\partial y^2}\right) u \\
  &\Leftrightarrow ((ik_x)^2+(ik_y)^2+(ik_z))\left(\frac{\hat{u}^{n+1}+2\hat{u}^n+\hat{u}^{n-1}}{4}\right) \\
  &= -(k_x^2+k_y^2+k_z^2)\left(\frac{\hat{u}^{n+1}+2\hat{u}^n+\hat{u}^{n-1}}{4}\right)
\end{align}
where $k_x, k_y,$ and $k_z$ are the wave numbers for the $x, y,$ and $z$ dimensions, respectively. In the 1D (and 2D) case, the $k_z$ (and $k_y$) wave numbers are not present.

We note that because the sine function is nonlinear, $\widehat{\sin u}\ne \sin\left(\hat{u}\right)$ and we must recompute $\widehat{\sin u}$ separately from $\hat u$ for every time step.

Putting everything together, we have
\begin{align}\label{eq:sgmethodunformatted}
&{}\frac{\hat{u}^{n+1} -2\hat{u}^n+\hat{u}^{n-1}}{\delta t^2}+ \left( k_x^2+k_y^2+k_z^2\right)\left( \frac{ \hat{u}^{n+1}+2\hat{u}^n+\hat{u}^{n-1}}{4}\right) \notag
\\&{} = -\widehat{\sin u^n}
\end{align}

Moving terms in \autoref{eq:sgmethodunformatted} around to compute $\hat{u}^{n+1}$ gives

\begin{equation}\label{eq:sgmethod}
  \begin{split}
    \hat{u}^{n+1} &= \frac{1}{\frac{1}{\delta t^2}+\frac{1}{4}\left(k_x^2+k_y^2+k_z^2\right)} \cdot \biggl[ \\
       & \frac{\left(2\hat{u}^n-\hat{u}^{n-1}\right)}{\delta t^2} -\frac{\left(k_x^2+k_y^2+k_z^2\right)\left(2\hat{u}^n +\hat{u}^{n-1}\right)}{4} - \widehat{\sin u^n}\biggr]
  \end{split}
\end{equation}

\section{Implementation}
\subsection{Simulation}
The 1D, 2D, and 3D codes are all implemented differently to illustrate many programming techniques. The 1D code is relatively straight-forward FORTRAN implementation. The 2D cases of interest are small enough to fit on one GPU, so the 2D code is implemented in C using Nvidia's CUDA. The FFTs and time-stepping scheme are run entirely on the GPU. We have found that GPU codes can run approximately 50 times faster for this type of code\cite{CloMuiRig12}. The 3D problems of interest use grid sizes that are too large to easily run on one computer, so we provide a parallel MPI implementation based on \cite{Tutorial12}. The 2decomp library is used to provide an efficient abstraction for performing 3D FFTs\cite{Ning10}.

\subsection{Visualization}
The 1D and 2D plots were generated using Python's Matplotlib\cite{Hu07}. The 3D plots were generated using VisIt\cite{Visit}. VisIt is useful for these because it can run in parallel. The 3D codes can generate a large amount of data, so this feature is very important.

\section{Results}
\subsection{1D Breathers}
\begin{figure*}[htb]
  \centering
  \subfigure[$t=0.6$]{\label{fig:u1d_start}
    \includegraphics[scale=0.25]{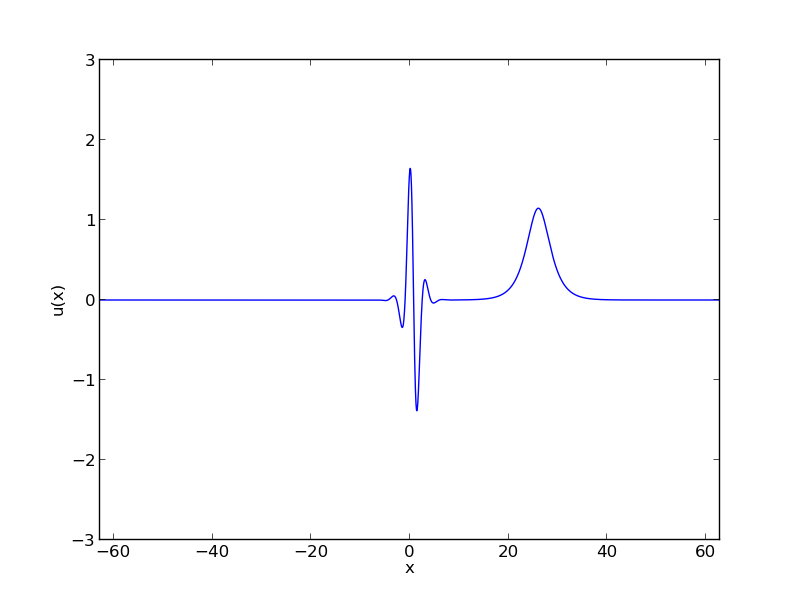}
  }
  \subfigure[$t=25.7$]{\label{fig:u1d_middle}
    \includegraphics[scale=0.25]{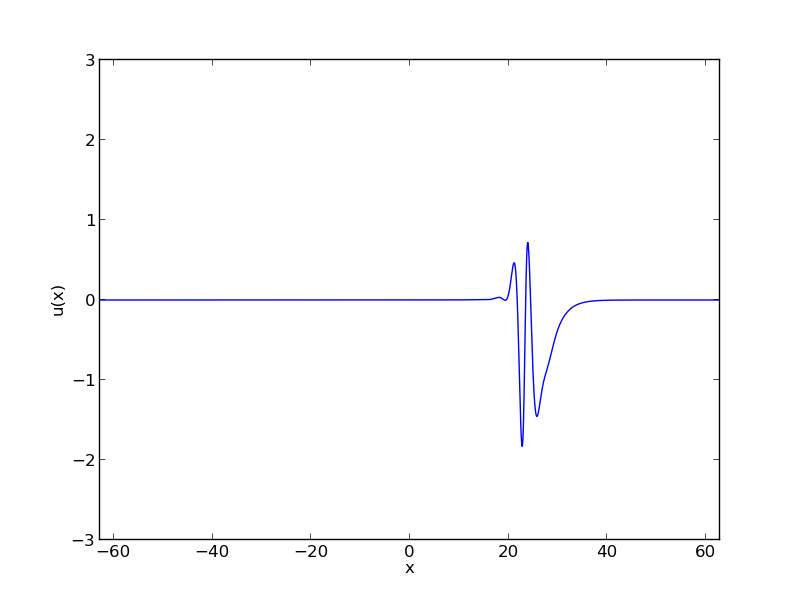}
  }
  \subfigure[$t=49.5$]{\label{fig:u1d_after}
    \includegraphics[scale=0.25]{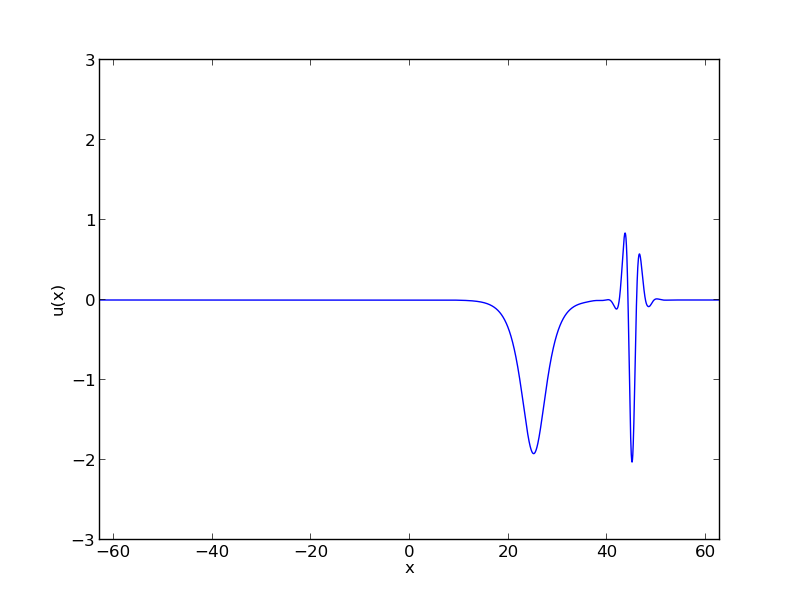}
  }
  \subfigure[Energy]{\label{fig:u1d_en}
    \includegraphics[scale=0.25]{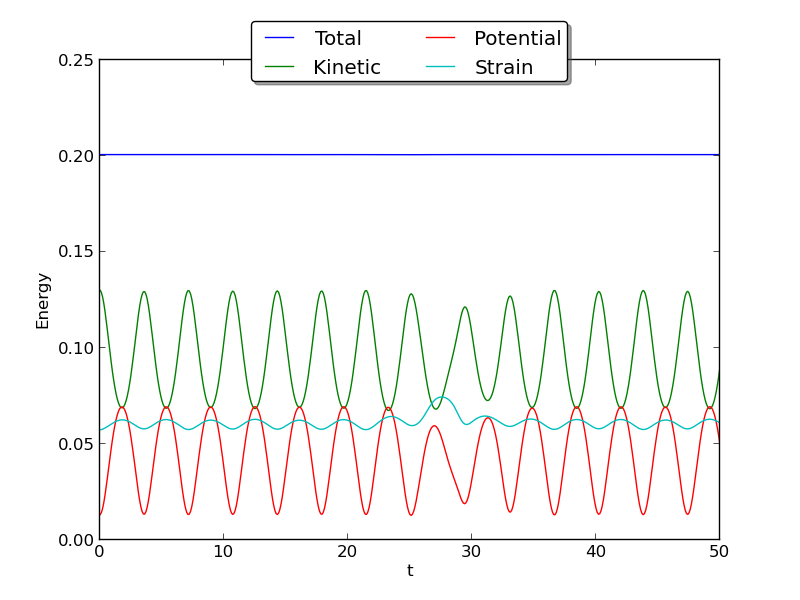}
  }
  \caption{\label{fig:u1d}The interaction between two breathers in one dimension. The middle breather is moving to the right and the breather on the right oscillates in place. \autoref{fig:u1d_start} shows before any interaction, \autoref{fig:u1d_middle} shows as the breathers are interacting, and \autoref{fig:u1d_after} shows after the breathers have interacted. \autoref{fig:u1d_en} shows how the various energies change with time. Note that energy oscillates between potential and kinetic and strain energy increases during the interaction between the two solitons, but the total energy remains constant. These were generated with $N_x=2^{12}, \delta t=0.05, \mu =2, c=0, 0.9$.}
\end{figure*}

One type of solution for the sine-Gordon equation is called a breather.
A breather is a nonlinear mode that is localized in space and oscillates with time. In \cite{En85}, Enz shows that these breathers can be interpreted as de Broglie waves; that is, a particle that is also a wave.
In the 1D case, a breather is given by\cite{SaScBiVa92}\cite{Wa96}
\begin{equation}\label{1dbreather}
u(x,t)=4\tan^{-1}\left( \tan\mu \frac{\cos \left( \gamma\cos\mu(t-xc)\right)}{\cosh \left( \gamma\sin\mu (x-ct) \right)} \right), \quad \gamma=\frac{1}{\sqrt{1-c^2}}
\end{equation}
where $\mu$ is a parameter that determines the size and frequency of the pulse and $c$ is the speed of the pulse.

\autoref{fig:u1d} shows a numerical solution with a stationary and moving breather colliding. After colliding, both breathers continue on their original trajectories.

\subsection{2D Breathers}
\cite{MiSmWo04} give initial conditions for stationary and moving breathers in 2D.

\begin{equation}
    u(x,y,t)  = -4\tan^{-1}\left[\frac{\lambda}{\sqrt{1-\lambda^2}}\sin\left( \phi(t)-kx\right)\sech\left( \lambda (t)(x-\zeta (t))\right)\sech \left( \lambda (t)y\right) \right]
\end{equation}

\begin{figure*}[htb]
  \centering
  \subfigure[$t=0.02$]{\label{fig:u2d_stat1}
    \includegraphics[scale=0.25]{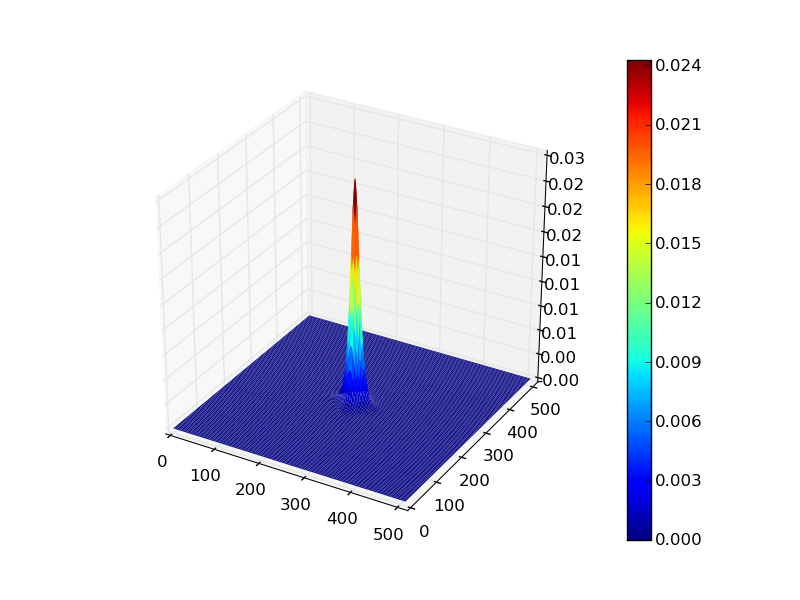}
  }
  \subfigure[$t=0.55$]{\label{fig:u2d_stat2}
    \includegraphics[scale=0.25]{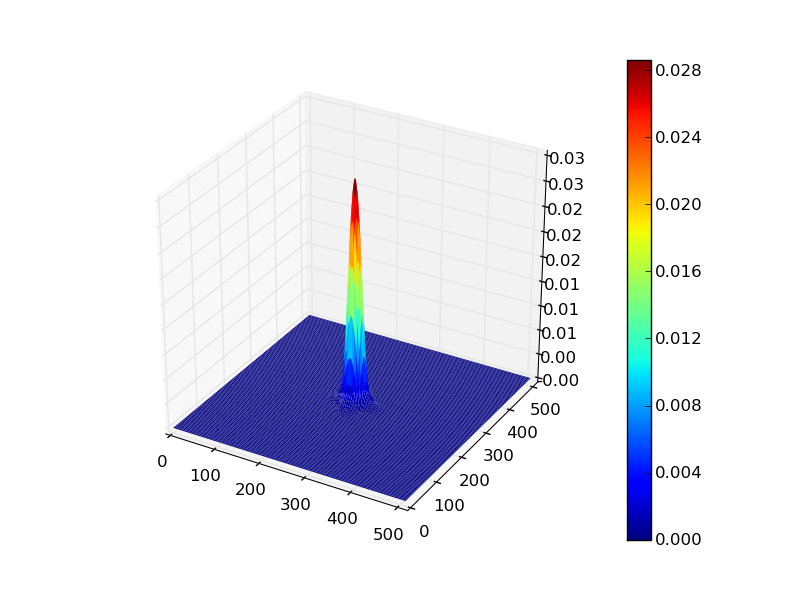}
  }
  \subfigure[$t=0$]{\label{fig:u2d_mov1}
    \includegraphics[scale=0.25]{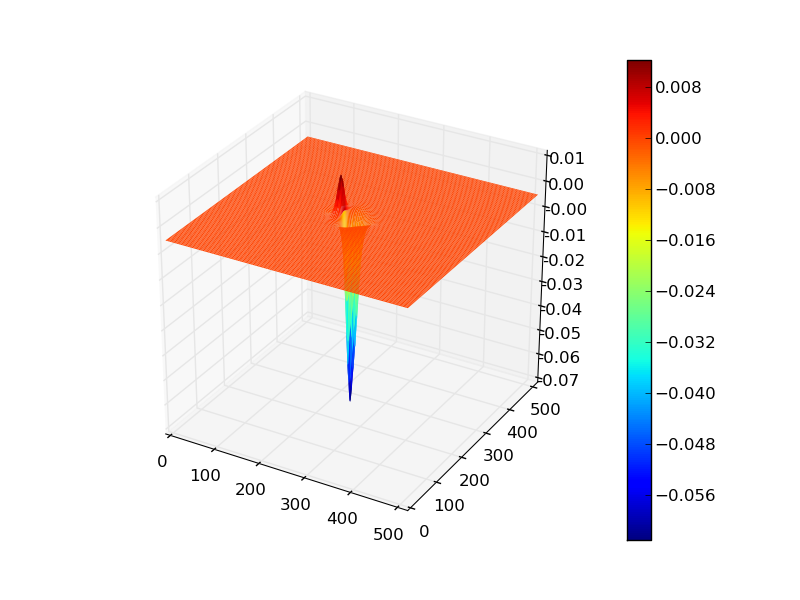}
  }
  \subfigure[$t=0.57$]{\label{fig:u2d_mov2}
    \includegraphics[scale=0.25]{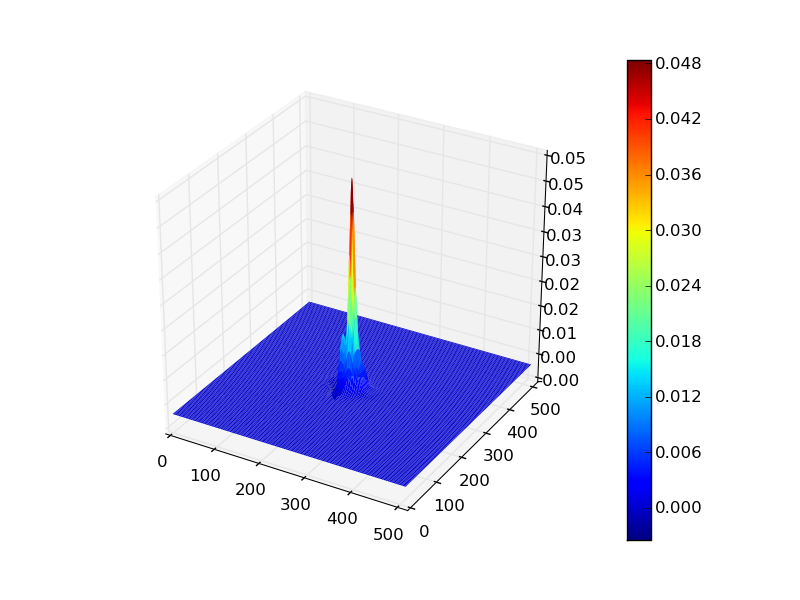}
  }
  \subfigure[Energy]{\label{fig:u2d_en1}
    \includegraphics[scale=0.25]{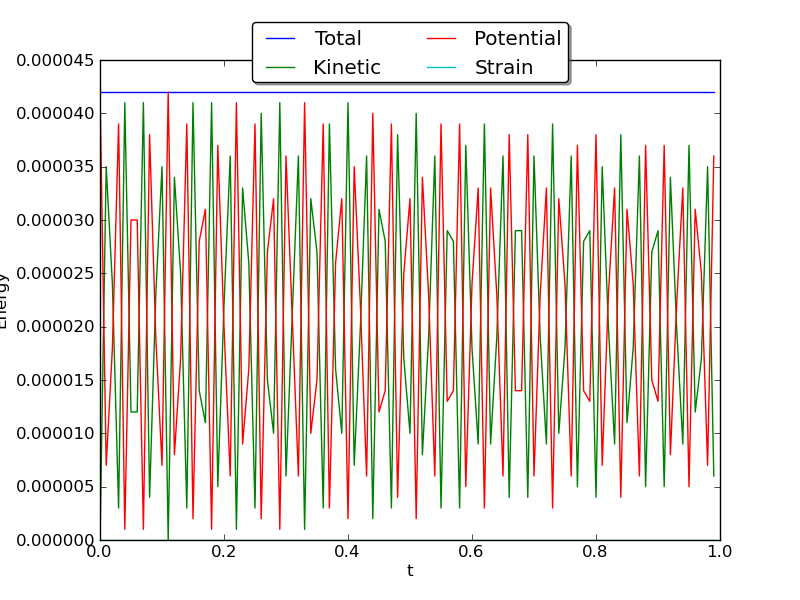}
  }
  \subfigure[Energy]{\label{fig:u2d_en2}
    \includegraphics[scale=0.25]{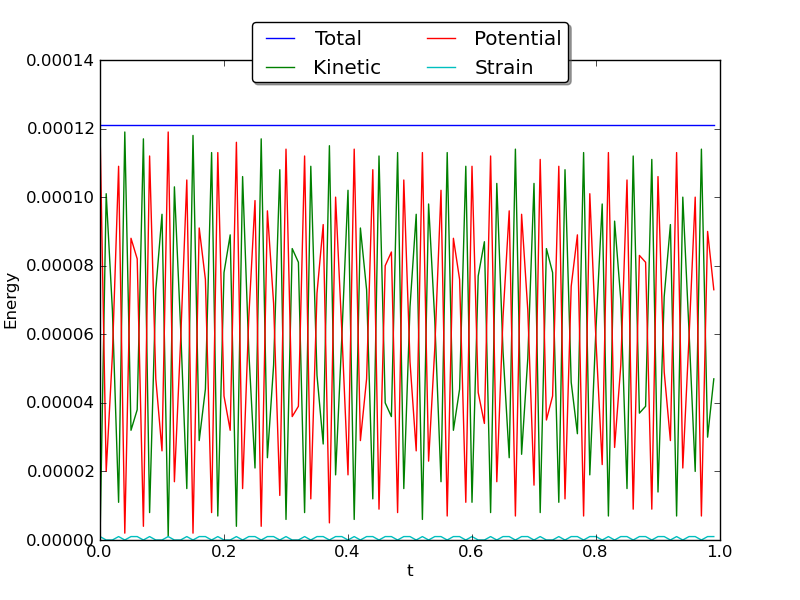}
  }
  \caption{\label{fig:u2d} \autoref{fig:u2d_stat1} and \autoref{fig:u2d_stat2} show a stationary breather at two different times. \autoref{fig:u2d_en1} shows the energy for this solution. \autoref{fig:u2d_mov1} and \autoref{fig:u2d_mov2} show a moving breather. \autoref{fig:u2d_en2} shows the energy for the moving breather.
 These were generated with $N_x=N_y=2^{9}, \delta t=0.01$. For the stationary breather, $\lambda (0)=0.1, \phi (0) =0.8, \zeta (0) = 0, k=0$. For the moving breather, $\lambda (0)=0.1, \phi (0) =0.8, \zeta (0) = 0.3, k=-0.1$.}
\end{figure*}

For $k=0$ and $\zeta (t)$ constant this gives a stationary breather, otherwise the breather is moving. \autoref{fig:u2d} shows the simulation results for initial conditions for a stationary and moving breather.

\autoref{fig:u2d} shows the results of running the simulation for the 2D ring solitons.

\subsection{3D Pseudo-Breather}
A 3D pseudo-breather was constructed based on \cite{dealII}. It is called a pseudo-breather because, like a breather,
it oscillates in place.
 The initial conditions are
\begin{equation}
\begin{split}
  u(x,y,z,t_0) = 4^3 &\tan^{-1}\left( \frac{\mu }{\sqrt{1-\mu ^2}}\sin (\sqrt{1-\mu ^2}t_0)\sech (\mu x)) \right) \\
  &\tan^{-1}\left( \frac{\mu }{\sqrt{1-\mu ^2}}\sin (\sqrt{1-\mu ^2}t_0)\sech (\mu y)) \right) \\
  &\tan^{-1}\left( \frac{\mu }{\sqrt{1-\mu ^2}}\sin (\sqrt{1-\mu ^2}t_0)\sech (\mu z)) \right)
\end{split}
\end{equation}

\begin{figure*}[htb]
  \centering
  \subfigure[$t=10$]{\label{fig:u3d_small}
    \includegraphics[scale=0.25]{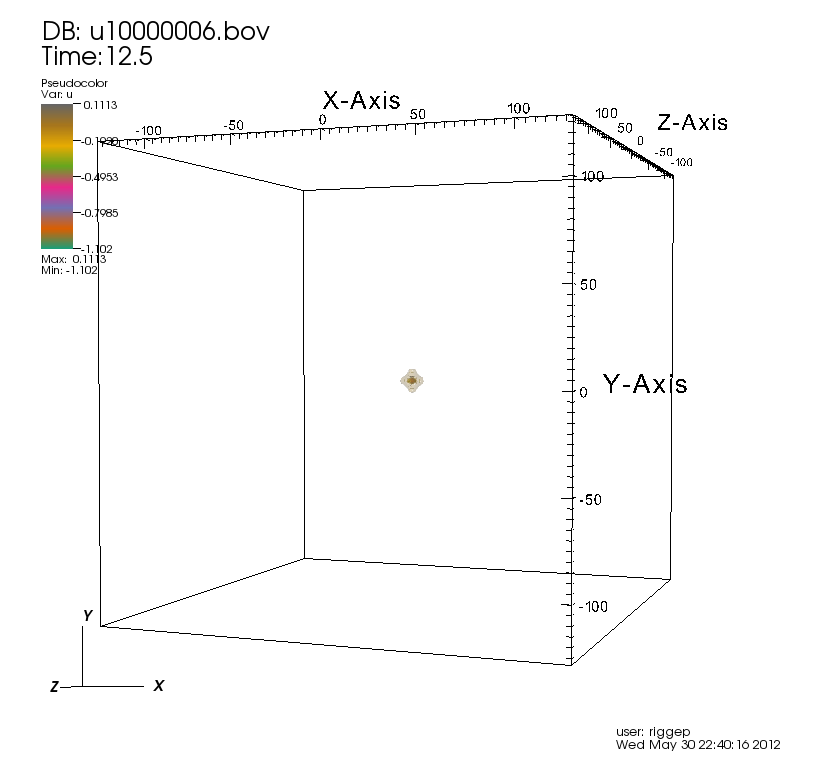}
  }
  \subfigure[$t=102.5$]{\label{fig:u3d_big}
    \includegraphics[scale=0.25]{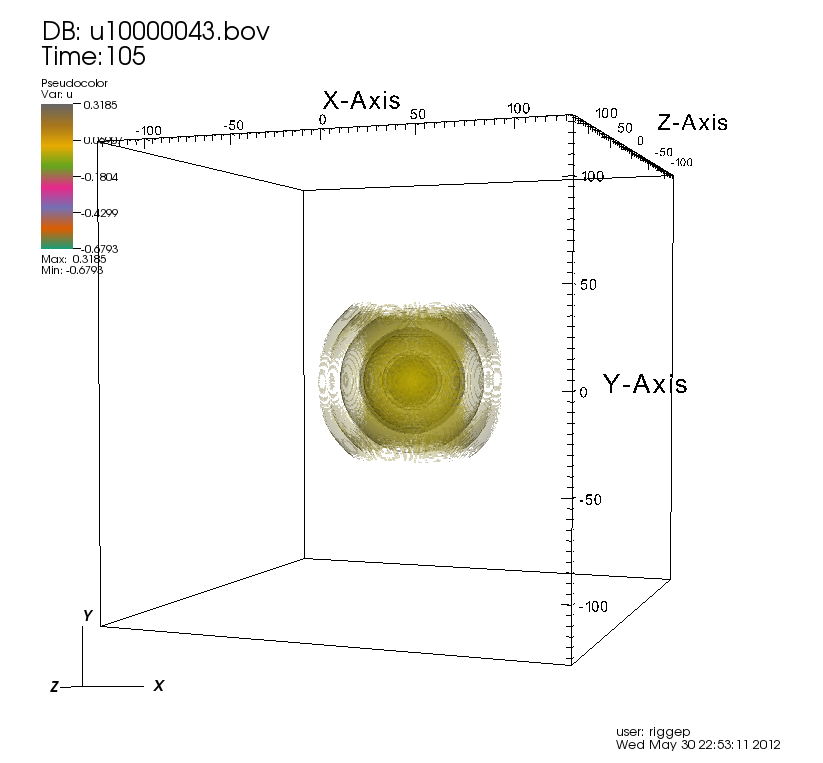}
  }
  \subfigure[Energy]{\label{fig:u3d_en}
    \includegraphics[scale=0.4]{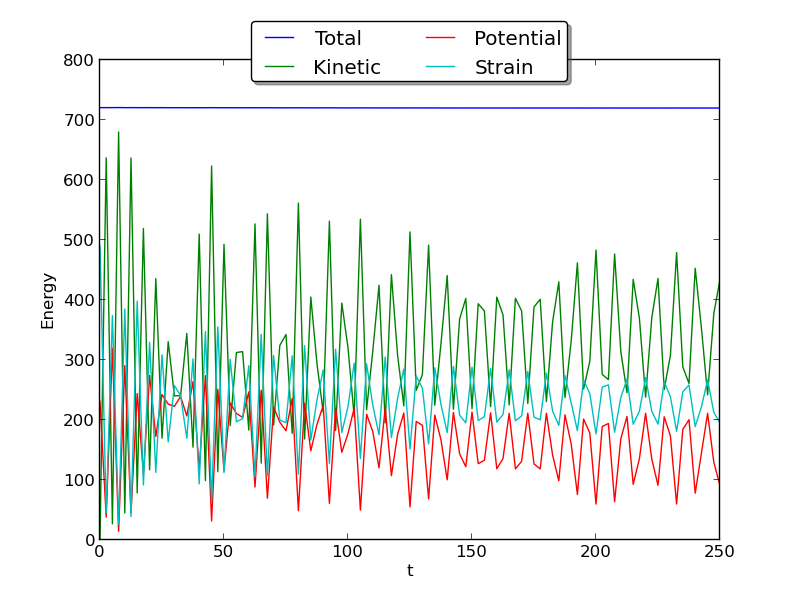}
  }
  \caption{\label{fig:u3d}A 3D pseudo-breather, shown with semi-transparent isosurfaces. The pseudo-breather oscillates in place, so the isosurfaces move in and out from the center of the pseudo-breather. \autoref{fig:u3d_small} and \autoref{fig:u3d_big} shows the pseudo-breather at different points between an oscillation. \autoref{fig:u3d_en} shows how the various energies change with time. These were generated with $N_x=N_y=N_z=2^{8}, \delta t=0.05, \mu =0.7, t_0=-5.4$. This was run on Flux with 8 cores.}
\end{figure*}

\clearpage

\section{Future Work}
So far, the 3D codes have only been run on relatively small machines with small grid sizes. Further work includes moving to a larger machine and larger grid sizes, as well as trying more complex interactions between a larger number of solitons.

\section{Acknowledgments}
This research was conducted as an internship with the Blue Waters Undergraduate Petascale Education Program (BW-UPEP). Many thanks to Shodor and the National Computational Science Institute (NCSI).

Many thanks to Benson Muite for his guidance, insight, and for giving me a great research experience.

I am also grateful to Brandon Cloutier for enjoyable and helpful discussions.

\vfill
\bibliographystyle{abbrv}
%%\bibliography{rigge}

\begin{thebibliography}{10}

\bibitem{dealII}
The step-25 tutorial program.
\newblock \\
  \url{http://www.dealii.org/developer/doxygen/deal.II/step_25.html}.

\bibitem{AsMc04}
U.~Ascher and R.~McLachlan.
\newblock On symplectic and multisymplectic schemes for the kdv equation.
\newblock {\em Journal of Scientific Computing}, 25(1):83--104, 2005.

\bibitem{Tutorial12}
G.~Chen, B.~Cloutier, N.~Li, B.~Muite, P.~Rigge, S.~Balakrishnan, A.~Souza, and
  J.~West.
\newblock Parallel spectral numerical methods.
\newblock 2012.

\bibitem{Visit}
H.~Childs, E.~Brugger, K.~Bonnell, J.~Meredith, M.~Miller, B.~Whitlock, and
  N.~Max.
\newblock A contract based system for large data visualization.
\newblock In {\em Visualization, 2005. VIS 05. IEEE}, pages 191--198. IEEE,
  2005.

\bibitem{CloMuiRig12}
B.~Cloutier, B.~Muite, and P.~Rigge.
\newblock A comparison of cpu and gpu performance for fourier pseudospectral
  simulations of the navier-stokes, cubic nonlinear schr\"odinger and sine
  gordon equations.
\newblock 2012.

\bibitem{CoTu65}
J.~Cooley and J.~Tukey.
\newblock An algorithm for the machine calculation of complex fourier series.
\newblock {\em Math. Comput}, 19(90):297--301, 1965.

\bibitem{DoSc10}
R.~Donninger and W.~Schlag.
\newblock Numerical study of the blowup/global existence dichotomy for the
  focusing cubic nonlinear klein-gordon equation.
\newblock 11 2010.

\bibitem{En85}
U.~Enz.
\newblock The sine-gordon breather as a moving oscillator in the sense of de
  broglie.
\newblock {\em Physica D Nonlinear Phenomena}, 17:116--119, 1985.

\bibitem{Ga67}
C.~S. Gardner, J.~M. Greene, M.~D. Kruskal, and R.~M. Miura.
\newblock Method for solving the korteweg-devries equation.
\newblock {\em Phys. Rev. Lett.}, 19:1095--1097, Nov 1967.

\bibitem{GrMa78}
G.~Grella and M.~Marinaro.
\newblock Special solutions of the sine-gordon equation in 2 + 1 dimensions.
\newblock {\em Lettere Al Nuovo Cimento (1971 -- 1985)}, 23:459--464, 1978.
\newblock 10.1007/BF02770537.

\bibitem{Hu07}
J.~Hunter.
\newblock Matplotlib: A 2d graphics environment.
\newblock {\em Computing in Science \& Engineering}, pages 90--95, 2007.

\bibitem{Ning10}
N.~Li and S.~Laizet.
\newblock 2decomp\&fft-a highly scalable 2d decomposition library and fft
  interface.

\bibitem{MaTl06}
P.~Mandel and M.~Tlidi.
\newblock Introduction to soliton theory (version 1.0).

\bibitem{MiSmWo04}
A.~Minzoni, N.~Smyth, and A.~Worthy.
\newblock Evolution of two-dimensional standing and travelling breather
  solutions for the sine--gordon equation.
\newblock {\em Physica D}, 189:167--187, 2004.

\bibitem{Ra01}
J.~Ramos.
\newblock The sine-gordon equation in the finite line.
\newblock {\em Applied Mathematics and Computation}, 124(45):93, 2001.

\bibitem{SaScBiVa92}
A.~S{\'a}nchez, R.~Scharf, A.~Bishop, L.~V{\'a}zquez, et~al.
\newblock Sine-gordon breathers on spatially periodic potentials.
\newblock {\em Physical Review A}, 45(8):6031--6037, 1992.

\bibitem{ScChMc73}
A.~C. Scott, F.~Y.~F. Chu, and D.~W. McLaughlin.
\newblock The soliton: A new concept in applied science.
\newblock {\em Proceedings of the IEEE}, 61(10):1443--1483, October 1973.

\bibitem{Tre00}
L.~N. Trefethen.
\newblock {\em Spectral Methods in Matlab}.
\newblock SIAM, 2000.

\bibitem{Wa96}
J.~Wattis.
\newblock Variational approximations to breathers in the discrete sine-gordon
  equation ii: moving breathers and peierls-nabarro energies.
\newblock {\em Nonlinearity}, 9:1583, 1996.

\end{thebibliography}
%%\include{rigge.tex.bbl}
%%\printbibliography

\end{document}